# Book Chapter-3

# A Survey on the Security and the Evolution of Osmotic and Catalytic Computing for 5G Networks

Gaurav Choudhary[1], Vishal Sharma[1*]

[1]Department of Information Security Engineering, Soonchunhyang University, Asan-si, South Korea, 31538, E-mail: gauravchoudhary7777@gmail.com, vishal_sharma2012@hotmail.com

**Abstract.** The 5G networks have the capability to provide high compatibility for the new applications, industries, and business models. These networks can tremendously improve the quality of life by enabling various use cases that require high data-rate, low latency, and continuous connectivity for applications pertaining to eHealth, automatic vehicles, smart cities, smart grid, and the Internet of Things (IoT). However, these applications need secure servicing as well as resource policing for effective network formations. There have been a lot of studies, which emphasized the security aspects of 5G networks while focusing only on the adaptability features of these networks. However, there is a gap in the literature which particularly needs to follow recent computing paradigms as alternative mechanisms for the enhancement of security. To cover this, a detailed description of the security for the 5G networks is presented in this article along with the discussions on the evolution of osmotic and catalytic computing based security modules. The taxonomy on the basis of security requirements is presented, which also includes the comparison of the existing state-of-the-art solutions. This article also provides a security model, "CATMOSIS", which idealizes the incorporation of security features on the basis of catalytic and osmotic computing in the 5G networks. Finally, various security challenges and open issues are discussed to emphasize the works to follow in this direction of research.

## 3.1. Introduction

The next generation of wireless networks exponentially adopts various emerging technologies to facilitate the service with a high-speed data rate [1]-[5]. The 5G technology is adopted in various applications like Industrial-Internet of Things (IIoT), smart city, and smart grid, to fulfill the user requirements and fasten the speed with low latency and high reliability for remotely accessing the regular services [6]-[8]. The 5G setup aims at exploiting different types of nodes which ensure on-demand as well as reliable connectivity to other devices [9]-[11]. With a tremendous increase in the number of devices, the pressure of maintaining the quality as well as the quantity will increase which also puts considerable impact on the security policies of a network.

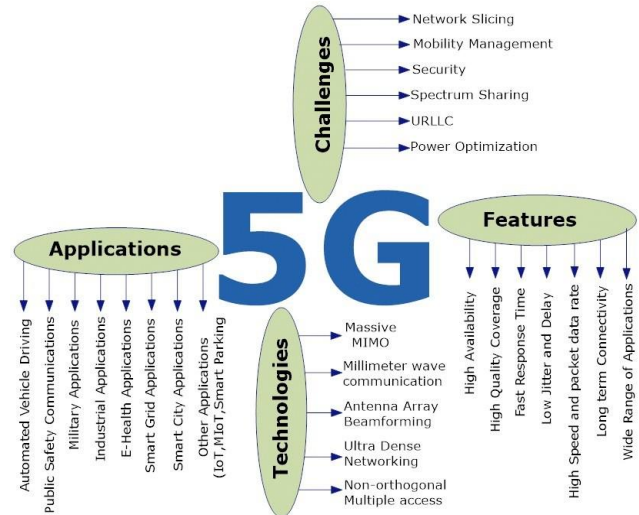

Fig 1. An illustration of 5G applications, technologies, features and challenges.

Different technology enablers such as blockchain, distributed mobility management, edge computing, osmotic computing, catalytic computing or fog networks can help to sustain this pressure imposed by the security requirements. However, there is gap in the literature as there are not sufficiently evident solutions which can cover both the performance as well as the security aspects of 5G networks. Use of diversified sensor nodes, drones, and autonomous vehicles can further impose extensive challenge on using 5G services without being compromised [6] [7] [10] [11]. Thus, with the rapid development of technologies in the 5G era, security becomes a major concern for successful implementations. Moreover, threats and vulnerabilities have been evolving continuously with the networks because of an increase in the number of users where attackers try to exploit the potential weaknesses. To easily follow these issues and aspects, an overview of 5G applications, technologies, features and performance challenges and standardization are presented in Fig.1 and Fig.2. Current 5G networks are supported by different technology enablers, which include Multiple-Input Multiple-Output (MIMO), Massive-MIMO, Non-Orthogonal Multiple Access (NOMA), Simultaneous Wireless Information and Power Transfer (SWIPT), Orthogonal Frequency Division Multiple Access (ODFMA), Radio Access Networks (RAN), Network Function Virtualizations (NFV), Software Defined Networks (SDNs), Device to Device Communications (D2D), Network Slicing, Low Power Wide Area Networks (LPWAN), etc [12]-[21].

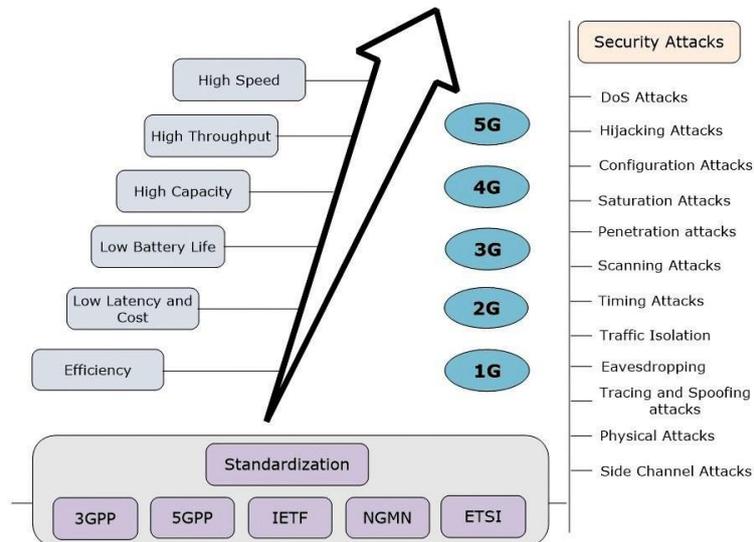

Fig 2. An illustration of 5G evolution, standardization and known security attacks.

### 3.5.1 **Applications of 5G networks:** The major applications of the 5G networks are provided below:

1. *Autonomous Vehicle*: The automatic controlled driving car and vehicles are key enablers of Vehicle-to-Vehicle (V2V), Vehicle-to-Infrastructure (V2I) and other Intelligent Transport Systems (ITS). The 5G network supports large bandwidth and low latency for these applications with high connection reliability. This network supports collision avoidance and intelligent navigation for the reliable transportation systems [6] [7].

2.   *Public Safety Communications (PSCs):* The PSCs facilitated various communication services in case of emergencies when the primary communication infrastructure is not available. The PSCs incorporated rapid deployment and accessibility of communication set-up. Therefore, 5G enabled communications supports a wide range of applications and long-term connectivity for the services in case of emergencies [12].

3.   *Military Applications*: The Military involved mission-critical control application which requires high data rate and long-term connectivity including the security parameters. The real-time surveillance and monitoring of suspected areas require a network with the large bandwidth and low latency. The 5G enabled network application is the best fit for such mission-oriented applications.

4.   *Industrial Applications*: Industrial automation composes massive IoT networks which require high connection density and low power consumption that can be ensured through 5G setups [22] - [24].

5.   *e-Health Applications*: The e-Health application requires remote diagnosis and long-term monitoring. The e-Health set-up involves video streaming embedded devices and advanced robotics which operate over the network that has Low power, low latency, and high throughput requirements. The 5G facilitates these requirements and serve as the best solution for these applications.

6.   *Smart City Applications:* Smart city adopts IoT devices, connected utilities, transportation, healthcare, education, smart grid etc. These applications scenario requires automation, cloud infrastructure and artificial intelligence which operate over a network composing large bandwidth, high throughput, high connection density and low latency [10]. The 5G networks can also be adopted by new technologies and industry enablers like robotics and drones, etc.

3.5.2   **Attacks and Threats in 5G networks:** Security holds a three-way relationship with the ease of deployment and user-friendliness. A more user-friendly setup often has poor security considerations and ease of deployment also reduces the level of security for any application. Such a tradeoff results in a large number of security attacks, which are the result of intentional or unintentional vulnerabilities. In most of the scenarios, adversaries take advantage of known vulnerabilities and launch attacks while exploiting them over a period of time. Some of the most hazardous attacks include zero-day attacks, self-exploitable attacks, or side channel attacks [3] [15]. Along with these attacks, there a certain set of known attacks, as shown in Fig.2, which impose a huge impact on the performance of any network. Considering the reachability and number of devices active in 5G networks, the scale and the impact of these attacks becomes severe and allows zero time to respond.

Efficient mutual authentication, strong key agreements should be followed while deploying applications in 5G networks, however, these should come without affecting the performance of the network [25]-[27]. Although 5G networks aim at increasing the speed as well as the quality of the transmissions, this reduces the time window for applying security mechanisms to prevent any active attacks on the networks. Thus, new mechanisms are needed for securing these fast communications or applicability approach of existing solutions needs to be changed to cope with the low-latency requirements of these networks [28]-[30]. Along with these attacks, it is desired the network should be free from insider threats as these may expose the internal functionalities which may lead to different types of attacks that are next to impossible to detect at the real time. Further, it is recommended to develop low-cost light-weight Intrusion Detection Systems (IDS), which can fixate on these requirements and can help identify attacks prior to their launching.

### 3.2. Preliminaries: Osmotic Computing

Osmotic computing is a network integration paradigm inspired by the chemical process of osmosis. In chemistry, the osmosis is a spontaneous mechanism of transferring or movement of the solvent through a semi-permeable membrane into solute concentrated solution. The solution is a mixture in which one substance is dissolved in another, the solute is the substance that is getting dissolved, and the solvent is the substance that is causing the dissolving. The standard mechanism of osmosis was molded with respective of available information by Villari et al. [31] to form this new paradigm of osmotic computing. This type of computing supports load balancing by enabling the movement of micro-services between a data center and the edge devices. It also reduces the latency of the overall application, however, in the initial drafts, the authors do not fully identify the mechanisms that can be used to support and facilitate such service migrations as stated in Sharma et al [32].

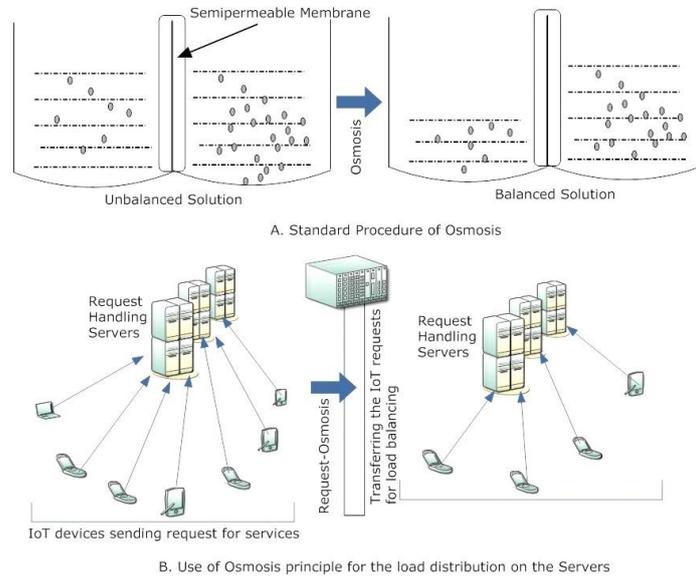

Fig 3. The standard osmosis procedure and the use of osmosis principle for load balancing on servers.

As per the standard mechanism of osmosis, the model contains a solution of the solute and the solvent and a semi-permeable membrane for filtering the solute. The participants of any model are interchanged with the respective terminologies to allow applicability of osmotic computing, like information and services, refer loads, process time, and energy of the system. The server can play the role of solvent for the interchangeable services through a semi-permeable membrane and is configured as a controller for the movement of services between the senders and the receivers. The basic principle of osmosis is used for load balancing of incoming requests. In case a server receives a lot of requests and is unable to handle in those instances, the service provider can act as a semi-permeable membrane and can shift the load to another server. To further enhance the understanding, the standard procedures without osmosis and load balancing with the osmosis principle are conceptualized in Fig. 3.

The primary objective of osmotic computing is to balance the load and resource utilization among servers without affecting the connectivity of services and performance. The real-time distribution of services as per the requirements helps to enable the new service migration concepts with the osmotic computing. Maksimović [33] defined the osmotic computing as "It enables the dynamic arrangement and migration of services and micro-services across cloud data centers and Edge resources according to differ-ent infrastructure demands and software." The author also conceptualizes the role of osmotic computing in the IoT and discusses the issues associated with the efficient execution of IoT services and applications across different computing infrastructures.

In the era of osmotic computing development, there have been limited but qualitative works that have highlighted the significance of using this new computing paradigm. Sharma et al. [32] focused on the efficient distribution and allocation of services via a concept of osmosis. The authors present a fitness-based Osmosis algorithm which is used to provide support for osmotic computing. The algorithm utilizes a fitness function to distribute and allocate the services into micro and macro-components. As per the authors, security concerns of the osmotic computing are still an open issue and future aspects should be considered with these issues. Furthermore, Sharma et al. [34] presented an efficient implementation of the osmotic computing, which is used for the pervasive trust management framework to perform computational offloading. The authors used three solutions which include the models- fitness-based movement, probabilistic movement, and threshold-based movement.

Nardelli et al. [35] presented the Osmotic Flow model, which supports multiple types and the mix of data transformation tasks on shared EDC+CDC infrastructures. The proposed model helps to assign the Edge and Cloud nodes among multiple and concurrent data transformation functions. As per the authors, the concept of machine learning with the distribution of data analysis tasks in various cloud environments imposes various solutions in a holistic manner. Morshed et al. [36] emphasized the deep learning concept in osmotic computing. The authors discussed the implementation issues and challenges of deep learning formations in the osmotic set-up.

Rausch et al. [37] emphasized the applicability of osmotic computing to the message-oriented middleware for enhancing reliability, ultra-low-latency, and privacy-aware message routing. The proposed solution enables the facilities for brokers to enable or diffuse the edge resources on the demand and support device to device communication in the IoT environment. Despite the facilities associated with osmotic computing, lots of challenges are also linked with the emerging technologies. Buzachis et al. [38] addressed the issues of connectivity in the environment which imposes network degradation and failures concerning osmotic nodes.

### 3.3. Preliminaries: Catalytic Computing

Catalytic computing is obtained from the principle of catalysis. In chemistry, catalysis involves the addition of a substance or material to move forward a reaction, where the added substance or material gets used by components of the chemical reaction. In the catalysis procedure, the substance which plays a significant role in the acceleration of chemical reactions is known as catalysts. The principal of catalysis is used with a network for providing efficient resource sharing without affecting the operations of other components and without concerning its own performance. As given in [39], the catalytic computing can be considered as a new paradigm to provide efficient resource sharing in wireless networks comprising users with high mobility. The authors used a Homogeneous Discrete Markov model for user mobility patterns to decide on the applicability of catalytic computing, selection of catalyst and procedures to fixate the activation energy. Furthermore, the authors present future aspect of newly coined paradigm for solving the problems like multi-network optimization problems, spectrum-sharing decision systems, scheduling problems, priority-based swarm communication, etc.

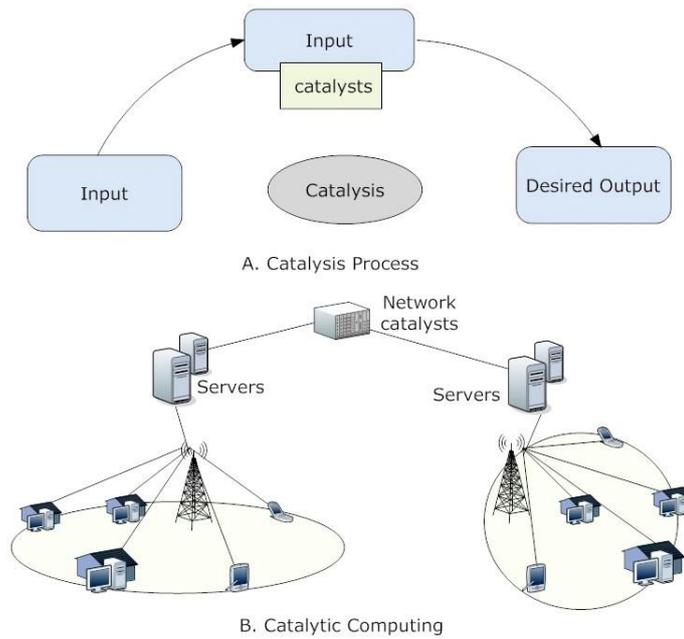

Fig 4. An exemplary illustration of the standard catalysis process and inspired catalytic computing for networks.

The standard process of catalysis and catalytic computing paradigm are illustrated in the Fig. 4. The catalyst in catalytic computing has the most important role to play and it operates between the entities allowing efficient management and utilization of network resources. The network cata-lysts are the supporting entities which facilitate the network activity and allow zero-downtime services to the users by acting as a decision system. The catalytic computing obeys the properties of the catalysis process and supports continuous decisions through centralized or distributed operations.

### 3.4. Existing Surveys and their Applicability

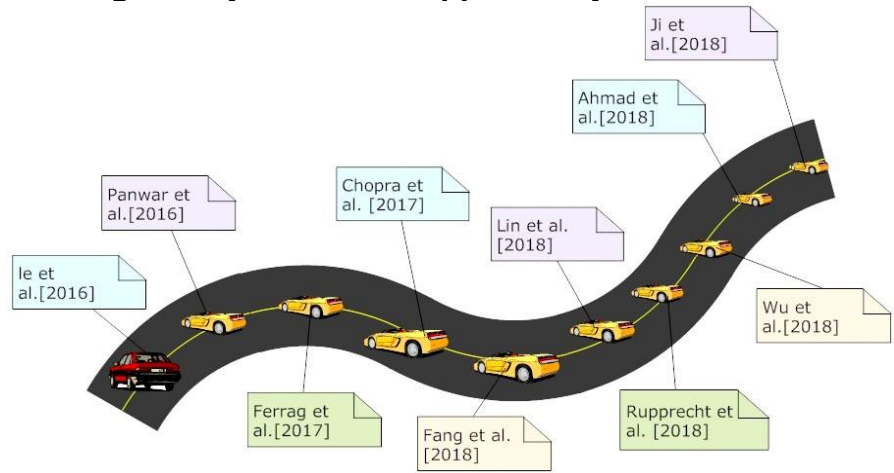

Fig 5. The roadmap of existing surveys on the security descriptions of 5G networks.

Over the last few years, various surveys have been published on the security issues and the challenges of 5G networks. Chopra et al. [40] discussed the security issues of the physical layer in Massive-MIMO, Jamming, Vehicular Ad Hoc Network (VANET), and D2D along with challenges and solutions for 5G ultra-dense wireless networks. The authors find the attacking regions in the ultra-dense networks. The authors also presented the types of jammers and their countermeasures. Ferrag et al. [41] discussed the authentication and privacy-preserving schemes in 5G. The authors presented the classification of the threat models and respective countermeasures against the attacks. Fang et al. [42] discussed the recent development of security solutions in 5G. The authors included a discussion on the security features with the recent technologies focusing Heterogeneous Networks (HetNet), D2D, Massive-MIMO, SDN, and IoT. The authors also paid attention to the 5G wireless security services like identity management and flexible authentications. Furthermore, the security services are also analyzed on the basis of the proposed architecture.

Lin et al. [43] presented the taxonomy of the network security-related data collection technologies. The authors analyzed the data collection node, mechanism, and tools. The authors also proposed the objective and requirements for the security-related data collections in a heterogeneous network. Furthermore, the research issues and future research challenges are considered in their survey. Rupprecht et al. [44] emphasized the main root cause of attacks and their defense in the mobile networks. The authors presented the challenges and research directions for the 5G security. The authors highlighted the limitations and drawbacks of existing works and security requirements for new 5G technologies.

The attacks and vulnerabilities of attacks, like DoS, are considerable issues for the future mobile generations. Ahmad et al. [45] presented the review of the challenges, and privacy and security issues in the clouds, software-defined networking, and network function virtualization along with their perspective security solutions. The major contribution of the authors includes the analysis of the attack-vulnerabilities and mitigation for making a secure 5G network.

The physical layer security approaches are robust and help to protect against eavesdroppers and man in the middle attacks. In addition, these provide flexibility for the secret key generation in 5G networks. Wu et al. [21] presented the survey on the physical layer security technologies Massive-MIMO, mmWAVE communications, heterogeneous networks, NOMA, full duplex technology, and challenges associated with these technologies. Furthermore, Gau et al. [46] gave a study of the physical layer security in the 5G based social networks. The authors discussed the challenges of the physical layer security in the 5G based large networks. The authors presented the solutions like security enhancement in the system layer, link layer, and cross-layer optimization.

Some more kinds of studies have been presented in Ji et al. [47] who analyze the security requirements of 5G business applications, network architecture, the air interface, and user privacy, and in Panwar et al.[48] who discussed the limitations of 4G networks and the features of 5G networks. The authors discussed the challenges, technologies, and implementation issues of these networks. Le et al. [49] presented the technology enablers for 5G communication and security challenges of the emerging technologies. From these studies, it can be marked that the security is a significant constraint for the 5G architecture and to facilitate the reliable services on the mobile communication it is necessary to consider privacy and trust of the user. The details of comparative evaluations and roadmap of existing surveys are presented in Table 1 and Fig.5.

Table 1. A comparative overview of the existing surveys on the security of 5G networks.

| Survey | Year | Application Area | Key Contribution |
|---|---|---|---|
| Chopra et al. [40] | 2017 | Ultra-dense network | Discuss security issues, challenges and respective solutions for 5G ultra dense wireless networks. |
| Ferrag et al.[41] | 2017 | Cellular networks | Presents the classification and comparisons of the authentication and privacy-preserving schemes, threat models and respective countermeasures against the attacks. |
| Fang et al. [42] | 2018 | 5G Mobile Wireless Networks | The authors discuss the recent development of security solution in the era of 5G. The author includes discussion on the security features with the recent technologies like HetNet, D2D, Massive-MIMO, SDN and IoT. |
| Lin et al. [43] | 2018 | Network Security | The authors present the taxonomy of the network security-related data collection technologies. The authors analyze the data collection nodes, mechanism and tools. |
| Rupprecht et al. [44] | 2018 | Mobile Network Generations | The authors discuss the root cause for attacks and their defense. The authors discuss the challenges and research directions on the 5G security. |
| Ahmad et al.[45] | 2018 | 5G Networks | The authors review the challenges and issues of privacy and security in the clouds, software defined networking, and network functions virtualization and presents respective solution for these issues. |
| Wu et al.[21] | 2018 | 5G Wireless Networks | The author presents the survey on the physical layer security techniques, challenges associated with these technologies and future trends. |
| Ji et al.[47] | 2018 | 5G Networks | The authors analyze the security requirements of 5G business applications, network architecture, the air interface, and user privacy. |
| Panwar et al.[48] | 2016 | 5G mobile communication | The authors discuss the limitations of 4G and features of 5G. The authors discuss challenges, technologies, proposed architecture and implementation issues. |
| Le et al.[49] | 2016 | 5G Networks | The authors discuss the technologies which enable 5G mobile communications. The authors emphasized on the challenging issues and future directions in 5G mobile access networks. |

## 3.5. Taxonomy of Security Concerns for 5G Networks

This section discusses the five major paradigms identified for the security of 5G networks as shown in Fig. 6. The section provides an overview of existing approaches along with their comparative evaluations.

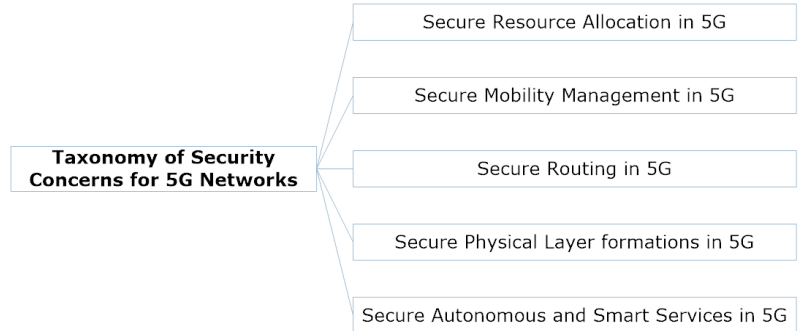

Fig 6. An illustration of classification of security concerns in 5G Networks.

### 3.5.1  Secure Resource Allocation in 5G:

In wireless communications, resource allocation plays a significant role. The resource allocation is defined as the management of resources and increases the capacity for a better communication. The various resources like power, load, spectrum, etc, are shared according to the user requirement for an optimized connectivity. The resources in the cellular networks can be assigned locally or globally. The traditional solutions emphasized the research allocations strategies by incorporating mathematical game theory problems but do not put a compatibility with security facilities. The security among resource sharing is essential from the perspective of optimization and better use of resources. The eavesdropper or an attacker can use the resources in a falsified way and increase the network degradation rate. Therefore, the secure resource allocation in the 5G networks is a crucial challenge and should be focused for improved connectivity.

Various studies have been presented in the era of secure resource allocation in 5G networks. Yu et al. [50] presented the optimal resource sharing scheme on the basis of the matrix game theory. The user adopts the concept of Enhanced Cloud-Radio Access Network (EC-RAN) for supporting data-heavy applications in the vehicular networks in a 5G communications environment. The authors also discussed the cloudlet resource management approach. Luong et al. [51] focused the economic and pricing approaches for resource management in the 5G networks. The authors use the resource management issues like user association, spectrum allocation, interference and power management to define appropriate models. Zhang et al. [52] emphasized the secure resource allocation for orthogonal frequency division multiple access. The mixed integer programming and non-convex models are used for solving the optimization problems. The au-thors proposed a system model on the concept of secrecy two-way relay WSNs without cooperative Jamming. Abedi et al. [53] presented the limited rate feedback scheme by formulating the optimization problem.

The spectrum sharing enables various unlicensed bands, including the Industrial, Scientific, and Medical (ISM) band, and visible light communication for the dedicated application by improving the utilization efficiency of the available spectrum. Akhtar et al. [54] proposed an SDN-based spectrum sharing technique. Liu et al. [55] discussed the 3-D resource allocation techniques. For the sum rate maximization, the authors combined the antenna selection and the user scheduling algorithms. Yang et al. [56] presented the random based radio resource allocation approach for D2D communications. In addition, network slicing can be a considerable solution for enhancing the capacity and flexibility of the networks in secure resource allocation and management schemes. Tremendous adoption of the 5G network with different application facilitates massive traffic and data on the networks [57]. The details of comparison between the state-of-the-art solutions for secure resource allocation in 5G Networks in presented in Table 2.

Table 2. State-of-the-art solutions for secure resource allocation in 5G Networks. (R1: Reliability, R2: Low Latency, R3: Security, R4: Resource Allocation/ Scheduling, R5: Key Management, R6: Osmotic Principles, R7: Catalytic Principles)

| Approach | Author (s) | Ideology | R1 | R2 | R3 | R4 | R5 | R6 | R7 |
|---|---|---|---|---|---|---|---|---|---|
| Optimal resource sharing | Yu et al.[50] 2016 | Matrix game theoretical approach | Yes | Yes | No | Yes | No | No | No |
| Economic and pricing approaches for resource management | Luong et al. [51] 2017 | Resource management issues like user association, spectrum allocation, interference and power management | Yes | Yes | Yes | Yes | No | No | No |
| Secure resource allocation | Zhang et al. [52] 2016 | Mixed integer programming problem | _ | _ | Yes | Yes | No | No | No |
| Limited rate feedback scheme | Abedi et al. [53] 2016 | Formulate as an optimization problem | _ | _ | Yes | Yes | No | No | No |
| HAS: harmonized SDN-enabled Approach | Akhtar et al. [54] 2016 | Centralized management based on distributed Inputs | _ | Yes | Yes | Yes | No | No | No |
| 5G Network Slice Broker | Samdanis et al. [58] 2016 | logically centralized monitoring and control entity | Yes | Yes | No | Yes | No | No | No |
| Radio resource allocation method for D2D communication | Yang et al. [56] 2016 | Random-based approach | No | No | No | Yes | No | No | No |
| 3-D Resource Allocation | Liu et al. [55] 2017 | Combine antenna selection and user scheduling | No | No | No | Yes | No | No | No |

| Techniques | | | | | | | | | |
|---|---|---|---|---|---|---|---|---|---|
| Architecture for network-slicing-based 5G systems | Zhang et al. [57] 2017 | Based on SDN and NFV technologies | Yes | Yes | _ | Yes | No | No | No |
| Resource allocation for secure communication | Wang et al. [59] 2018 | Three-dimensional stochastic model | Yes | _ | Yes | Yes | No | No | No |
| Joint resource allocation framework | Li et al. [60] 2017 | heuristic VNE algorithms | No | Yes | _ | Yes | No | No | No |
| Green resource allocation scheme | AlQerm and Shihada [61] 2018 | Centralized and decentralized Sophisticated online learning scheme | No | No | No | No | No | No | No |
| Secure communications in NOMA system | Zhang et al. [62] 2018 | Formulate in the non-convex optimization Problem | No | No | Yes | Yes | No | No | No |
| Hierarchical radio resource management scheme | Belikaidis et al. [63] 2018 | spectrum access system | _ | No | No | Yes | No | No | No |
| Network resource allocation system | Martin et al. [64] 2018 | Machine Learning methods | Yes | Yes | No | Yes | No | No | No |
| Social trust scheme | He et al. [65] 2018 | deep learning approach | _ | _ | Yes | Yes | No | No | No |
| framework of resource allocation and performance optimization | Tan [66] 2017 | Based on control theory | Yes | Yes | Yes | Yes | No | No | No |
| Network slicing management & prioritization | Jiang et al. [67] 2016 | heuristic-based admission control mechanism | No | No | No | Yes | No | No | No |
| Energy efficient resource allocation | AlQerm and Shihada[68] 2017 | Machine Learning Scheme | No | No | No | Yes | No | No | No |
| SDN based resource Management | Duan [69] 2017 | Programmable management platform | _ | Yes | Yes | Yes | _ | No | No |
| Resource allocation scheme | Zhu[70] 2017 | geometric approach that analyzes the feasible region governed by the constraints | _ | _ | Yes | Yes | _ | No | No |
| Resource-based mobility management | Sharma et al. [39] 2017 | Catalytic computing | No | Yes | No | Yes | No | No | Yes |
| Service Management scheme | Sharma et al. [32] 2018 | Osmotic computing | No | Yes | No | Yes | No | Yes | No |

## 3.5.2  Secure Mobility Management in 5G:

The 5G network imposes a high data rate transmission in the communication and delivery of services. The high data rates can be achieved through better traffic control and mobility management. Various technologies in-corporated handover as a key concept for mobility management. For the better communication in the 5G scenario, various researches have been conducted for mobility management with the security consideration of the communications. In the era of mobility management in 5G, Sharma et al. [39] presented a resource based mobility management scheme. The proposed solution is based on the catalytic computing. The authors used a Homogeneous Discrete Markov model for analyzing mobility patterns of the user and congestion control algorithm. The handover mechanism is presented on the basis of activation energy.

Furthermore, for the handover schemes for the 5G networks, Sharma et al. [71] presented a secure key exchange and authentication protocol for fast handover in the 5G Xhaul networks. The proposed protocol provides security over the links for the moving terminals in the networks. The authors discussed the security of the Backhaul, fronthaul and Xhaul. You and Lee [72] gave a ticket based handover approach for the 5G networks. The proposed mechanisms rely on the concept of reducing messages involved in the authentication procedure with the authentication server. The user anonymity in the handover is considered by Fan et al.[73] in their ReHand scheme. Their scheme reduces the communication cost compared with the existing solutions.

Munir et al. [75] presented a secure fault tolerance mechanism for 5G networks based on the concept of the distributed hash table of access nodes and ticket-reuse approach. The authors presented an analytical model for analyzing the processing time of a location query with the minimum authentication latency. To support Distributed Mobility Management (DMM), Sharma et al. [76] developed a Blockchain-based DMM handover scheme. Their solution focused on the resolution of hierarchical security issues without affecting the network layout. Some other solutions can be studied from Refs [77-80]. In short, 5G should be adopted with a reliable and scalable architecture for the mobility management to support a high data rate with the minimum delays and latency. The details of comparison between the state-of-the-art solutions for secure mobility management in 5G Networks in presented in Table 3.

Table 3. State-of-the-art solutions for secure mobility management in 5G
Networks. (R1: Handovers, R2: Key Management, R3 Mutual Authenti-
cation; R4: Osmotic s/Catalytic Principles)

| Approach | Concept | Parameters Used | R1 | R2 | R3 | R4 |
|---|---|---|---|---|---|---|
| Key exchange and authentica-tion protocol [71] | Securing Xhaul links for a moving terminal in the network | Failure factor, Packet loss, Sig-naling overheads | Yes | Yes | Yes | No |
| SPFP : Ticket-based secure handover [72] | Decrease the number of message related to authen-tication | Handover latency, packet loss/buffering, and handover failure | Yes | Yes | Yes | No |
| ReHand: secure region-based handover scheme [73] | Incorporate the techniques of group key, one-time identity, and one-way hash | Computational cost, communi-cation cost | Yes | Yes | Yes | No |
| User-centric Ul-tra-dense net-working [74] | Dynamic AP grouping | Solutions for mobility man-agement and Mobility scenario | Yes | Yes | Yes | No |
| Secure and fault-tolerant mecha-nism for DMM [75] | Used distributed hash table of access nodes and ticket-reuse approach | Average authentication latency, average processing time | Yes | Yes | Yes | No |
| Blockchain-based DMM [76] | Resolving hierarchical se-curity issues without affect-ing the network layout | Energy consumption | Yes | Yes | Yes | No |
| Mobility and traf-fic management mechanism for delay tolerant cloud data [77] | Selecting appropriate cell for data transfer in down-link, and by configuring the UE cell priorities for send-ing the data in uplink | Throughput distribution, delay | No | No | No | No |
| Policy-based communications [78] | System controlled by policy to detect misbehavior | Session setup delays | No | No | No | No |
| Policy-based per- | The decisions are made on | Handover decision making | Yes | No | No | No |

| flow mobility management system [79] | the basis of policies. | | | | | |
|---|---|---|---|---|---|---|
| Secure producer mobility management [80] | Use concept of hash-chain that protects against prefix hijacking attacks occurring during mobility updates | Storage cost, edge router good-put, Verification delay | Yes | Yes | _ | No |
| Resource-based mobility management [39] | Catalytic computing | Equilibrium evaluation, Activation energy | Yes | No | _ | Yes |

### 3.5.3  Secure Routing in 5G:

The 5G networks enhance the connectivity of various applications with the high speed, low latency, and high availability. The standard architecture considers various network management challenges. One of the major considerations in these challenges is the secure routing. With the increase in traffic, it is very difficult to manage services and data in a real-time along with their secure deployments. The standard network management mechanisms pay attention to the routing and monitoring of data flow. The data transmission and routing should be protected against various kinds of attacks like the man in the middle, downgrading attacks, location spoofing, and resource depletion attacks.

For the efficient and reliable data communication in the 5G era, secure routing is an emerging research area. Over the years, an extensive research has been conducted on the secure routing in the 5G networks. Guan et al. [81] presented a GRBC-based network security functions placement scheme with applicability to the software-defined security. The authors focused on the loop-free routing schemes and independent routing decisions. Sharma et al. [39] discussed the routing –based mobility management, SDN-based mobility management and Cluster-based mobility management mechanisms. The routing based mobility management is used in the older concept with the benefit of the optimal path for traffic flow. The proposed approach, based on the catalytic computing, supports the user movement control and routing-based handovers.

Jung et al. [82] presented a joint operation protocol for the group key management and routing. Routing control helps to maintain the change in state of the links. Wang and Yan [83] discussed various secure routing schemes like the Secure Message Delivery (SMD) protocol, Secure Optimized Link State Routing protocol (SOLSR) that have a functionality to protect the messages relayed between the source and the destination. The authors also analyzed the secure routing, access control, and Physical Layer security requirements for D2D communications.

Choyi et al. [84] focused on the hybrid routing through network slicing. The authors also discussed the static and the dynamic mechanisms of routing. The static and the functional overview of the service negotiation framework and flexible routing with network slicing are also discussed. In their framework, routing is used by the static path defined by packets. Naqvi et al. [85] presented the study for the acceptance possibilities of the IPv6 within 5G wireless networks. The authors presented a study of inter-domain routing and the comparison study with the IPv4.

Schmittner et al. [86] focused the PSCs and presented a novel secure multi-ti-hop D2D scheme. The proposed scheme is used for forwarding decision that is based on reliability metric for the routing. Zhao et al. [87] presented a cluster-based technique focusing security based transmission protocol for Ultra-Dense Networks (UDNs). The authors analyzed the security issues and resource management schemes for better network formations. Furthermore, a secure routing protocol is presented for the multi-hop networks. The protocol is based on the concept of Weil Paring [88]. The secure routing enables a better communication and service management via traffic flow and optimized use of resources. Therefore, the secure routing mechanism should be included in the 5G network for a reliable and secure communications. The details of comparison between the state-of-the-art solutions for secure routing in 5G Networks in presented in Table 4.

Table 4. State-of-the-art solutions for secure routing in 5G Networks.
(R1: Encryption, R2: Mutual Authentication; R3: Key Management, R4: Osmotic s/Catalytic Principles)

| Approach | Mechanism | Parameters Used | R1 | R2 | R3 | R4 |
|----------|-----------|-----------------|----|----|----|----|
| Catalysis-based mobility management [39] | Homogeneous discrete Markov model for user mobility patterns | Equilibrium evaluation, Activation energy | No | No | No | Yes |
| Group Routing Betweenenss | Finding a group of given size with maximum GRBC | Complexity analysis, Network topology and Attack model | No | No | No | No |

| Centrality [81] | | | | | | |
|---|---|---|---|---|---|---|
| Joint operation protocol [82] | Use group key agreement method | Computational load, Latency time | Yes | No | Yes | No |
| Secure Routing [83] | Secure Message Delivery (SMD) protocol | D2D security architecture and security requirements | Yes | Yes | Yes | No. |
| Network Slices and routing [84] | Hybrid routing | Packet Routing | No | No | No | No |
| IPv6 technology [85] | OPNET MIPV6 model | Throughput, network delay and packet delay | No | No | No | No |
| SEMUD [86] | Per-destination routing state used | Throughput, Goodput | No | No | No | No |
| Cluster-based UDNs [87] | Security based Transmission Protocol | Network Eavesdropping Defense and Jamming Attack Defense | Yes | No | No | No |
| Secure routing protocol [88] | Weil Paring based | End-to-end delay, Throughput | Yes | _ | Yes | No |

### 3.5.4  Secure Physical Layer formations in 5G:

The physical layer security gains a hike in the 5G network communications. The conventional cryptographic solutions for security impose the difficulties in the distribution and management of secret keys. But the physical layer has provided reliable and flexible security levels through various protocols. The concept of the physical layer is to "utilize the intrinsic randomness of the transmission channel to guarantee the security in physical layer" [21]. The existing approaches to security require strong constraints and high additional costs for the users of public networks. Therefore the new concept on physical layer security focuses on the secrecy capacity of the propagation channel, which is adopted by the 5G networks.

Gomez et al. [90] analyzed the physical layer security for uplink NOMA in 5G large area networks. The key mechanism of coverage probability and the effective secrecy throughput is used in their solution. The Signal-to-Interference plus Noise Ratio (SINR) is measured with the legitimate user and attackers. Furthermore, Forouzesh et al. [91] analyzed the physical layer security for Power Domain (PD)-NOMA in HetNet. The novel concept of resource allocation scheme is presented to maximize the sum secrecy rate in PD-NOMA based HetNet. The interference in the signals is considered as an eavesdropper in their approach. An alternative search method algorithm is adopted for the optimization problem. Furthermore, Liu et al. [95] discussed the physical layer security for 5G-NOMA on the basis of stochastic geometry approaches and the location of the user. Asymptotic secrecy outage probability is used for the multiple-antenna and the simulation result is verified with the Monte Carlo mechanisms.

Massive-MIMO has the capability to handle 100+ antenna elements and a large number of independent transceiver chains. The increasing adaptability of Massive-MIMO opens the feasibility of adaption to physical layer security against the eavesdroppers. Kapetanovic et al. [92] presented the opportunities and the challenges of physical layer security for Massive-MIMO. The authors presented various possible attack methods and detection of the active attacks on the Massive-MIMO. Wang et al. [93] discussed the physical layer security in cellular HetNet. The authors proposed a threshold-based secrecy mobile association policy and calculated secrecy and connection probabilities of the random users. The closed-form expressions are used with the secrecy and connection probabilities.

Zhang et al. [96] presented the cooperative anti-eavesdropping techniques on the basis of graph theory. In their approach, on the basis of network topology, a secrecy weighted graph is calculated. The proposed techniques facilitate the low complex security on the large-scale networks. Chen et al. [97] analyzed the physical layer security for cooperative NOMA systems. Pan et al. [98] presented hierarchal physical layer security architecture for the 5G Networks based on the cross-layer lightweight authentication scheme and a physical layer security assisted encryption scheme. These schemes are based on the key streams and the channel information. An overview of the state-of-the-art solutions for the physical layer security-technologies in the 5G secure communications is presented in Table 5.

Table 5. State-of-the-art solutions for secure physical layer formations in 5G Networks. (R1: Technology Used, R2: Secrecy, R3: Channel Coding; R4: Jamming; R5: Osmotic/Catalytic framework)

| Approach | Mechanism | R1 | R2 | R3 | R4 | R5 |
|---|---|---|---|---|---|---|
| Physical layer se- | Coverage probability and the effective secrecy | 5G, NOMA | Yes | No | No | No |

| curity for NOMA [90] | throughput | | | | | |
|---|---|---|---|---|---|---|
| Physical layer security for PD-NOMA [91] | Used Alternative Search Method algorithm | 5G, NOMA | Yes | No | No | No |
| Physical layer security for Massive-MIMO [92] | On the basis of pilot contamination scheme | MaMIMO, 5G | Yes | No | Yes | No |
| Physical layer security in HCN [93] | On the basis of access threshold-based secrecy mobile association policy | 5G | Yes | No | No | No |
| Physical layer security for 5G-NOMA [94] | On the basis of stochastic geometry approaches, the location of user obtained. | 5G, NOMA | Yes | No | No | No |
| Physical layer security for NOMA [95] | Added eavesdropper exclusion area (protected area) | 5G, NOMA | Yes | No | No | No |
| Cooperative anti-eavesdropping techniques [96] | On the basis of graph theory | P2P, D2D, 5G | Yes | No | No | No |
| Physical Layer Security for Cooperative NOMA Systems [97] | Secrecy outage probability (SOP) and strictly positive secrecy capacity (SPSC) | 5G, NOMA | Yes | No | No | No |
| Hierarchical security architecture [98] | On the basis of cross-layer light weight authentication scheme | eMBB, mMTC and URLLC,5G | No | No | No | No |
| Robust beam forming [99] | On the basis of robust information and artificial noise (AN) beam forming | Massive-MIMO, 5G, BDMA | No | No | No | No |
| Hybrid MIMO phased-array time modulated directional modulation scheme [100] | Time-modulated DM scheme is added for the phased-MIMO to achieve PLS | MIMO, 5G | Yes | No | No | No |

### 3.5.5  Secure Autonomous and Smart Services in 5G:

Various architectures for autonomous services have been discussed that support the flexibility, performance, cost, security, safety, manageability, etc [101]. Cheng et al. [102] discussed the architecture of 5G-based IIoT in the three application modes; namely, enhance Mobile Broadband (eMBB), massive Machine Type Communication (mMTC), Ultra-Reliable and Low Latency Communication (URLLC). The authors presented the applicability of the IIoT and Cyber-Physical Manufacturing Systems (CPMS) in the 5G wireless communication network and also discussed the key challenges in the adaptability of these networks.

Mavromatis et al. [103] presented architecture for connected and autonomous vehicles. This architecture relies on the mechanism of multi-layer application data streaming. The authors presented the next-generation ITS for intelligent traffic planning, smart emergency vehicle routing, and multimodal commuting. Han et al. [105] discussed the mobile sensing and cloud computing and presented a combined concept of mobile cloud sensing.

The low latency applications in 5G networks have emerged stronger and are adopted widely for the business perspective. Lema et al. [106] discussed the applications that rely on the low latency in the area of healthcare, automotive and transport systems, entertainment, and manufacturing from the market perspectives and respective models. Saghezchi et al. [107] emphasized the smart grid for smart cities and presented secure network architecture on the basis of detection of price integrity or load alteration attacks. Furthermore, Arfaoui et al. [108] presented the security architecture for 5G networks. From these studies, it is conclusive that the adaptable architecture for autonomous services in the 5G networks should provide flexible solutions and benefits for industries and leverage the new technology enablers for the overall development, especially security features, as a perspective of reducing the research and deployment cost. The details of comparison between the state-of-the-art solutions for secure services in 5G Networks in presented in Table 6.

Table 6. State-of-the-art solutions for secure services in 5G Networks.
(R1: Technology Used, R2: Security, R3:  IDS Formations; R4: Business Models; R5: Osmotic/Catalytic framework)

| Approach/ Architecture | Mechanism | R1 | R2 | R3 | R4 | R5 |
|---|---|---|---|---|---|---|
| 5G wireless communication technology[102] | On the basis of technical scheme and features of 5G wireless communication technology | 5G, eMBB, mMTC and URLLC | Yes | No | _ | No |

| Architecture for connected and autonomous vehicles [103] | Multi-layer application data streaming | 5G, Multiple Radio Access Technologies | No | No | Mobility-as-a-Service (MaaS) | No |
|---|---|---|---|---|---|---|
| 5G Network Architecture [104] | Intelligent use of network data | 5G, Massive-MIMO, NFV, and SDN. | Yes | No | IaaS and XaaS | No |
| Architecture of mobile cloud sensing [105] | Mobile sensing and cloud computing | 5G, Opinion Finder, Google Profile of Mood States (GPOMS) | No | No | all-IP network (AIPN) model | No |
| Use case with ultra-low latency in 5G [106] | Market perspectives of each industry and respective models. | 5G, Advanced imaging, data analysis, and machine learning | Yes | No | B2B model | No |
| Secure network architecture for smart grids [107] | On the basis of detection of price integrity or load alteration attacks | 5G, AMI, M2M | Yes | Yes | - | No |
| Security architecture for 5G networks [108] | Inherited concepts from the security architectures of 3G and 4G networks | 5G, ITU-T X.805, 3GPP | Yes | No | Trust Model | No |
| Fog Computing [109] | Based on the TSI- NFV MANO architecture | 5G, Peer-to-Peer (P2P) | Yes | Yes | _ | No |
| Architecture for monitoring services. [110] | Self-protection mechanism through monitoring | 5G, SDN, NFV | Yes | Yes | Business support systems | No |
| Fog-based anomaly detection approach [111] | Unsupervised Clustering and Outlier detection algorithms | 5G, LPWAN, Edge computing, SDN, NFV | Yes | No | _ | No |

### 3.6. CATMOSIS: A generalized model for 5G Security

Both the osmotic and the catalytic computing can be united together to form an efficient security module which can be used by all types of applications. One such example is shown in Fig. 7. The figure shows four different catalytic managers for four different types of applications each of which are handled through the osmotic servers for cross-platform exchange of information. The intra mode security of each application can be obtained through traditional mechanisms, however, with the amalgamation of osmotic and catalytic computing, it becomes convenient to secure the cross or inter-platform services. This generalized model operates in two parts to form a CATMOSIS module for security as given below:

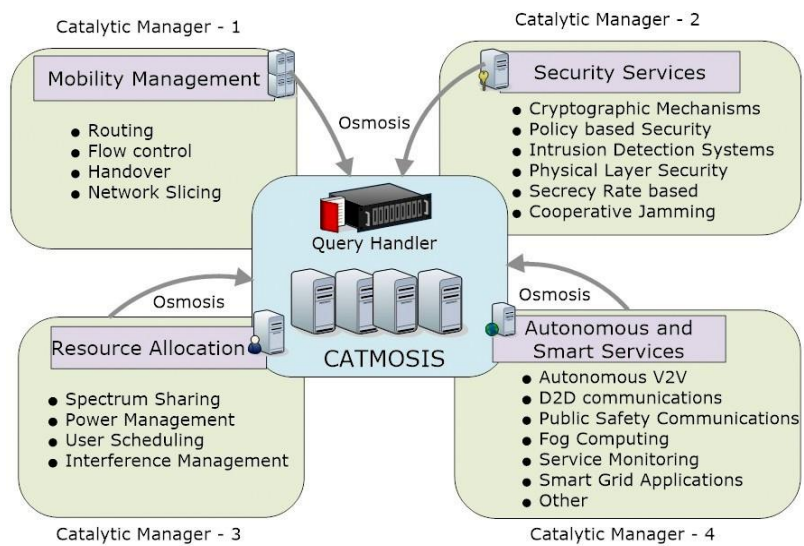

Fig.7. An illustration of the CATMOSIS general model for security enhancement of 5G networks.

- **Catalytic Manager:** The module is responsible for handling all the intra-mode services and generate dependencies for secure communications. This module is similar to the one proposed by the original authors. It uses activation energy based resources to decide on the policies and security requirements which are to be shared over the network platform.
- **Osmotic Manager:** This module is responsible for fetching the services to and from the catalytic manager while keeping intact the general flow of traffic. It helps to distinguish the security requests from the general traffic and allows categorization feature for providing enhanced security. However, the security is enhanced depending on the protocol or the methodology opted for authentication of devices as well as services. At the moment, flow authentication and categorization can be attained through this module and other possibilities of extending this are left to future works.

Table 7: An overview of other studies to follow for the security of the 5G networks.

| Focus | Author (s) | Challenges and Issues | Solutions |
|---|---|---|---|
| Analysis of the physical downlink and uplink control channels and signals. | Lichtman et al. 2018 [112] | Jamming Vulnerability, Sniffing and Spoofing Vulnerability | Mitigation techniques |
| 5G Security | Schneider and Horn 2015 [113] | Flexible Security and potential security requirements | Security mechanisms for 5G- like User Identity and Device Identity Confidentiality, Mutual Authentication and Key Agreements, Security between Terminal and Network and protection against attacks |
| 5G Mobile Communication Networks | Luo et al. 2015 [114] | Traditional mechanisms does not find the potential vulnerabilities in the 5 G-SDN-MN | vulnerability assessment mechanism |
| 5G Vehicular Networks | Ejiyeh and Talouki 2017 [115] | Attacks-Replay, message fabrication and DoS attacks | Efficient security protocol |
| Network security in mobile 5G | Bouras et al. 2017[116] | SDN security challenges: Eavesdropping, Identity spoofing, Password-related attacks | General security guidelines, Firewalls, administrative passwords, Testing Techniques etc. |
| Security for 5G Communications | Mantas et al.[117] | Potential threats and attacks for the following 5G system components: the UE, the access networks, the mobile operator's score network and the external IP networks. | Potential mitigation schemes |
| 5G security | Svensson et al.[118] | Information leakage between network slices | Architecture for authentication, authorization and accounting for 5G |
| 5G wireless communication network | Gupta et al. 2017[119] | Bandwidth spoofing attack | Attack modeling and intrusion detection system |
| 5G Security | Ahmad et al. 2017[120] | Security Threat, security issue in SDN and NFV and in communication channels. | Security Technology- DoS, DDoS detection, link verification, access controls, identity and location security etc. |
| Network Slicing in 5G | NGMN Alliance [121] | Issues: Controlling Inter-Network Slices Communications, Impersonation attacks, variation in the security protocols and resources exhaustion. | Mutual authentication, integrity verification, baseline security levels for each slices etc. |

| 5G Technologies | Felita and Suryanegara 2013 [122] | Discuss the technological challenges like security, Limited frequency spectrum resources. | Suggests: TRIESTE framework, BDMA technique |
|---|---|---|---|
| 5G Wireless Communication Networks | Yang et al. 2015 [123] | Focus on the security challenges in various technologies: HetNet, Massive-MIMO and mmWave. | Physical layer security |
| 5G Networks | Sun and Du 2017[124] | Challenges for Network Security | Physical layer Security approaches- Artificial noise injection, Anti-eavesdropping signal Design, Secure beam forming /precoding etc. |
| Cloud computing in heterogeneous 5G | Gai et al. 2016 [125] | crucial security concerns in mobile cloud computing | Intrusion detection techniques |
| 5G cellular networks | Atat et al. 2017[126] | cellular transmissions protection and links protection from eavesdropping | Suggest security controls (authentication), monitoring of real-time data streams, implementing advanced anomaly detection techniques, etc |
| 5G Network Security | Mammela et al. 2016 [127] | security visibility and configurability issues, threats and vulnerabilities issues | implementation of security monitoring spanning multiple micro-segments |
| 5G wireless communication networks | Gandotra and Jha 2017[128] | Secure power optimization issues | Standardization agencies and discuss on the ongoing projects |

## 3.7. Open Issues and Future Directions

5G era is focusing on the enhancement of applications which otherwise are difficult to attain at current network functionalities. Some of the other works to follow for enhancing the understandings on the 5G and its security are discussed in Table 7. Features like virtual reality networks, connectivity to billions of IoT devices and information on the go are some of the major enablers in 5G setup [129]. However, with applications gaining new heights, security becomes primary as well as tedious objective to attain [130]. Thus, following these requirements, this article summarizes major security directions to be followed for further research as provided below:

1.    ***Secure and Flexible Architecture:*** The standard architecture should be formed to support generalized applications from different domains. The flexibility in terms of adoption of various add-on features with the compatibility of the application perspectives is needed in the security architectures [131].  The security should be considered in different do-mains to handle attacks and security threats. Various secure architectures have been suggested to support the domain-specific application with the facilities of reliability and efficiency [132]-[134]. But from the research perspective, the adopted architecture should be enabled to follow generalized implementation mechanism within the application domains and sustain the security within the architecture.

2. ***Secure and continuous connectivity***: The massive increase in IoT devices in the 5G networks raises various connectivity issues. The connection should be continuous and stream less to facilitate real-time monitoring and service scheduling. The 5G network should facilitate security policies without affecting the services and connectivity of the network while maintaining the privacy of its users [135].

3. ***Sustainability and reliability:*** The application dependability is measured in terms of the reliability and the sustainability of the network. The resource depletion and unwanted service consumes a lot of resources over the network and impose an extra load on the network which results in sudden failures. The sustainability should be a considerable issue to resolve along with the efficiency and the reliability. The 5G based applications require a wide range of the devices to connect with high connection density and low power consumption, which require a focus on the reliability and the sustainability to enhance the network tolerance level [136] [137].

4. ***Strong and efficient Mutual authentications:*** The security mechanisms in 5G should incorporate strong mutual authentication to verify the originality of both the receiver and the sender. The authentication mechanisms can be performed in various phenomenons over the network such as handover, mobility management, and D2D communications [138][139]. Therefore, a strong and efficient mutual authentication mechanism is required which maintains the freshness of keys and session while securing the applications in the 5G setup.

5. ***Secure spectrum sharing and network slicing***: Network slicing is a key concept of resource allocation and management. The interference in the network imposes signal distortion and noise. To reduce such kinds of issues, optimized and secure spectrum sharing should be adopted to enable large applications with different user priorities. Such a requirement can be attained through SDN-NFV technologies while leveraging the properties of network slicing [140].

## 3.8. Conclusions

This paper provided a detailed description of the security for the 5G networks. The evolution of osmotic and catalytic computing and the details of their implementation and utilization are also presented in this article. The roadmap of different kinds of attacks and their possible solutions are highlighted in the initial parts of this article. Furthermore, the taxonomy on the basis of security requirements is presented, which also includes the comparison of the existing state-of-the-art solutions. In addition, existing surveys, open issues, and security challenges are discussed to provide a research direction. Moreover, this article also provides a security model, "CATMOSIS", which idealizes the incorporation of security features on the basis of catalytic and osmotic computing in the 5G networks.